\documentclass[conference]{IEEEtran}
\IEEEoverridecommandlockouts

\usepackage{graphicx}
\graphicspath{{Myfig/}}
\usepackage{float}
\usepackage{algorithm}
\usepackage{algpseudocode}
\algtext*{EndIf}
\algtext*{EndFor}
\usepackage{amsmath}

\usepackage{cite}
\usepackage{amsmath,amssymb,amsfonts}
\usepackage{textcomp}
\usepackage{xcolor}
\def\BibTeX{{\rm B\kern-.05em{\sc i\kern-.025em b}\kern-.08em
    T\kern-.1667em\lower.7ex\hbox{E}\kern-.125emX}}
\begin{document}
\title{Maintaining Data Integrity in Fog Computing Based Critical Infrastructure Systems}

\author{\IEEEauthorblockN{ Abdulwahab Alazeb, Brajendra Panda}
\IEEEauthorblockA{\textit{Dept. of Computer Science and Computer Engineering} \\
\textit{University of Arkansas}\\
Fayetteville AR, USA \\
afalazeb,bpanda@uark.edu}

}

\maketitle

\begin{abstract} 
The evolution of the utilization of technologies in nearly all aspects of life has produced an enormous amount of data essential in a smart city. Therefore, maximizing the benefits of technologies such as cloud computing, fog computing, and the Internet of things is important to manage and manipulate data in smart cities. However, certain types of data are sensitive and risky and may be infiltrated by malicious attacks. As a result, such data may be corrupted, thereby causing concern. The damage inflicted by an attacker on a set of data can spread through an entire database. Valid transactions that have read corrupted data can update other data items based on the values read. In this study, we introduce a unique model that uses fog computing in smart cities to manage utility service companies and consumer data. We also propose a novel technique to assess damage to data caused by an attack. Thus, original data can be recovered, and a database can be returned to its consistent state as no attacking has occurred. 

\end{abstract}


\section{Introduction}
The future of the internet is in the Internet of Things (IoT), evidenced by the significant increase in wearable technology, smart homes and buildings, connected vehicles, and smart grids. The number of connected IoT devices in use by 2025 is estimated at nearly 75 billion \cite{rr12}, producing an enormous amount of data predicted to total more than 79 zettabytes \cite{rr13}. Limitations and restrictions in bandwidth, as well as this rapid growth in the amount of data produced, make the current information system architecture inadequate for managing and moving that volume of data to the cloud. In many scenarios, with increasing use of IoT devices, it would be impractical to do so. 
Additionally, contemporary society has incorporated a staggering number of sensitive and real-time IoT applications as integral parts of daily life.  Connected car technologies, video conference applications, health monitoring, and real-time production line monitoring, are all applications requiring low-latency and location awareness in order to provide high-quality services for this technology-driven era \cite{r2}.

IoT devices, such as smart meters in modern smart cities, will not only produce a massive amount of data, but heterogeneous data that will a need to be processed in real-time \cite{rr4}. This data will not be valuable enough without exploiting the maximum benefits from other technologies. It is unmanageable and nearly impossible for a cloud to handle the many tasks, of processing and aggregating, analyzing and storing for such volumes of disparate data \cite{rr2}. For that reason, fog computing was presented by Cisco \cite{r3}. 

Fog computing is trending today because it has several unique characteristics that, not only establish it as an extension of the cloud, but also provide privileges in addition to, and complementing. those of the cloud. This complement to cloud computing, in an age of cloud processing, is significant in allowing the performance of resource, intensive, and extended term analytics \cite{rr14}. 

In the potential for smart cities, fog computing will help the IoT and smart meter devices process the data and make quick decisions to take the right action within a critical time-frame and to aggregate only the indispensable data for the cloud. Services and utility companies, such as water and electric companies, can exploit fog computing technology to manage and analyze the volume of consumers’ data. There are many existing studies on how to expand the efficiency of fog computing in smart cities and to solve technical issues related to the large volume of data that needs fusion and integration to the cloud \cite{rr15}. Moreover, while security and privacy issues have been addressed by many researchers, other aspects need more attention, such as a case in which a protection system fails during a cyberattack and costumers’ data need to be recovered. This study aims to detect all transactions that are affected by any malicious transaction, recover the correct value of data, and ensure the integrity of consumer data in a fog computing environment in smart cities.

\section{Literature Review}
As a new network technology, fog computing has attracted many researchers. Quite a number of papers have presented a general survey or study of the challenges, issues, and future direction of fog computing \cite{rr10}. Other works write about the architecture of the fog system \cite{rr5}. Security and privacy issues are also a hot topic in this field \cite{rr3, rr11}.

Many researchers have written about the use of fog computing in smart grids and cities to improve the quality of services provided to consumers and to solve security and privacy matters. Zhu et al. \cite{rr1} proposed a new architecture for using fog computing in smart cities. They presented a privacy enhancement scheme that would use techniques such as blind signature to provide anonymous authentication on consumer data. They also proposed applying extra encryption techniques on the smart meter readings of all consumer data at times. This can be done when each fog node is aggregated to the cloud. Although Zhu et al. claimed that their proposed scheme would enhance the privacy and security of consumer data, their model is still vulnerable to electronic and insider attacks.

Lyu et al. \cite{rr3} proposed a new framework to securely aggregate smart meter readings through fog nodes and to the cloud. To ensure consumer data privacy, their data will be encrypted after adding some statistical noise using the Gaussian noise technique. Furthermore, Aazam et al. \cite{rr5} discussed the architecture of the Industrial Internet of Things, which is the use of the IoT in the manufacturing industry for such applications as smart sensors, actuators, and robots. It is important for fog computing to be a solution that provides essential support closer to end users to ensure local and real-time processing for sensitive and complex tasks.

Much research has been conducted in the area of damage assessment and data recovery. Researchers have proposed models and mechanisms that try to recover data after cyberattacks. The column dependency based approach introduced by Chakraborty et al. \cite{rr6}, observes the relationship between transactions to determine which transactions have been affected by malicious attacks and need to be recovered. In this way the time-consuming recovery of data after attacks will take less time than traditional approaches. Chakraborty et al. proposed a recovery method that will take the affected transactions as input and perform the recovery in two stages: compensation and re-execution. Their experiments indicated that when malicious transactions increase in the database, the second stage of their recovery scheme is also increased.

Liu \cite{rr7} aimed to improve the efficiency of damage assessment and repair in distributed database systems. They first identified the challenges and complications that those systems face and then proposed an algorithm for distributed damage assessment and repair. On each site, they adopted a Local Damage Assessment and Recovery (DAR) Executor to scan the local log to detect and clean any sub-transactions that were affected by a malicious transaction. Also, on each site there is a Local DAR Manager that cooperates with the Executor to ensure global coordination between all sites on the system by generating a coordinator for any cleaning transaction.

Panda et al. \cite{rr8} used the data dependency-based approach to assess the damage that could occur from electronic attacks and then to return the database to a consistent state. They introduced two algorithms. In the first one, damage assessment and recovery algorithms are executed at the same time, blocking the system until the whole procedure is complete and causing significant delays as a result. The second algorithm addresses the delays since the system should be available soon after all the affected and damaged data have been detected and blocked.

Alazeb and Panda \cite{rr11} introduced two different models for using fog computing in a healthcare environment. The first model is an architecture that uses fog modules with heterogeneous data, and the second model uses fog modules with homogeneous data. For each model, they propose a unique methodology to assess the damage caused by malicious transactions so that original data may be recovered and affected transactions identified for future investigation.
\section{Model}
In this section, a unique architecture for using a distributed fog node system in smart cities to manage the consumer data of utility services will be proposed. Then, cooperative algorithms will be proffered for identifying, assessing, recovering, and restoring all the damaged and affected data created by an attack. The goal is the restoration of a reliable database. In the proposed model, it is assumed that the Intrusion Detection System (IDS) is responsible for detecting malicious transactions in the system and providing a list of those transactions to the damage assessment algorithms. Each fog node in the proposed architecture must have its own log file and use a strict serializable history. All operations in the log file need to be in the same order in the history.
The log files cannot be user modified at any time. Since the log files will contain a record of every modification to the value of any data item that is updated by write operations, all read operations are also required to be stored in the log files to identify the data dependencies between the operations and the transactions and among the detections of the victim fog nodes in the systems.

\subsection{Model Nations}
A description of the notations to be used in the proposed model can be found in Table~\ref{table1}. 
\begin{table}[htbp]
\caption{Notation used in our proposed approach description}
\begin{center}
\begin{tabular}{|l|p{2.4in}|}
\hline
\textbf{Notation}&\textbf{Description}\\
\hline
pub\_fog & The public fog nodes that are accessed by customers and utilities providers.  \\
\hline
usc\_fog & The private fog node for each utility service company.  \\
\hline
MT\_L & The list of detected malicious transactions done by IDS.  \\
\hline
DA\_Table & The damage audit table, which is a data structure that will be created by the damage assessment algorithms to collect data about transactions that are needed to do the data recovery, such as the valid and invalid read data items, data written, and accessed fogs.     \\
\hline
DI\_L & The damage item list that will contain all damaged data items that are identified by our proposed damage assessment algorithms. \\
\hline
DIT\_ Fog$_x$ & The Fog$_x$ damage item table, where x is the ID of the secondary affected fog node, which reads any damage data item from another fog node.   \\
\hline
VIT\_ Fog$_x$ & The valid data items table that will be created by algorithm 3 or 4 to add to it all recovered data items for the secondary affected fog node Fog$_x$. It will be sent to Fog$_x$ to use it as input on algorithm 4.\\
\hline
$w_i(A, v_1, v_2)$ & The write operation of the transaction T$_i$; $v_1$ is the before image, which represents the old value of the data item A before any updating. And $v_2$ is the after image, which is the new value of data item A after it is updated. \\
\hline
$r_i(A, v)$ & The read operation of transaction T$_i$ where A is the data item and v is the current value of A.   \\
\hline
\end{tabular}
\label{table1}
\end{center}
\vspace{-0.2 in}
\end{table}
 

\subsection {The Proposed Architecture}
In the proposed model, each smart city will have several public fog nodes (pub\_fog), which will be efficiently distributed to guarantee the quality of service at each point of the entire city. Private fog nodes will be included, with at least one private fog node for utility service companies (usc\_fog), such as water, electricity, and gas utilities. The usc\_fog nodes should be effectively located in the center of the whole distributed system to ensure a reliable connection to all pub fog nodes and provide different routes should one of the pub\_fog nodes disconnect for any reason. Consumers will be able to send queries to pub\_fog nodes only. Data may be retrieved from the local database if available there. Otherwise, the queries will be forwarded to the appropriate usc\_fog node. Consumers are not allowed to directly connect the usc\_fog nodes for security reasons. All queries related to those nodes will come through the pub\_fog nodes. Customer utility usage data will be collected from smart homes and buildings using IoT devices and smart meters. Usage data will be sent to the nearest efficient pub\_fog nodes based on several factors, such as location and load balance. 
\begin{figure} [!tbp]
\includegraphics[width=\columnwidth]{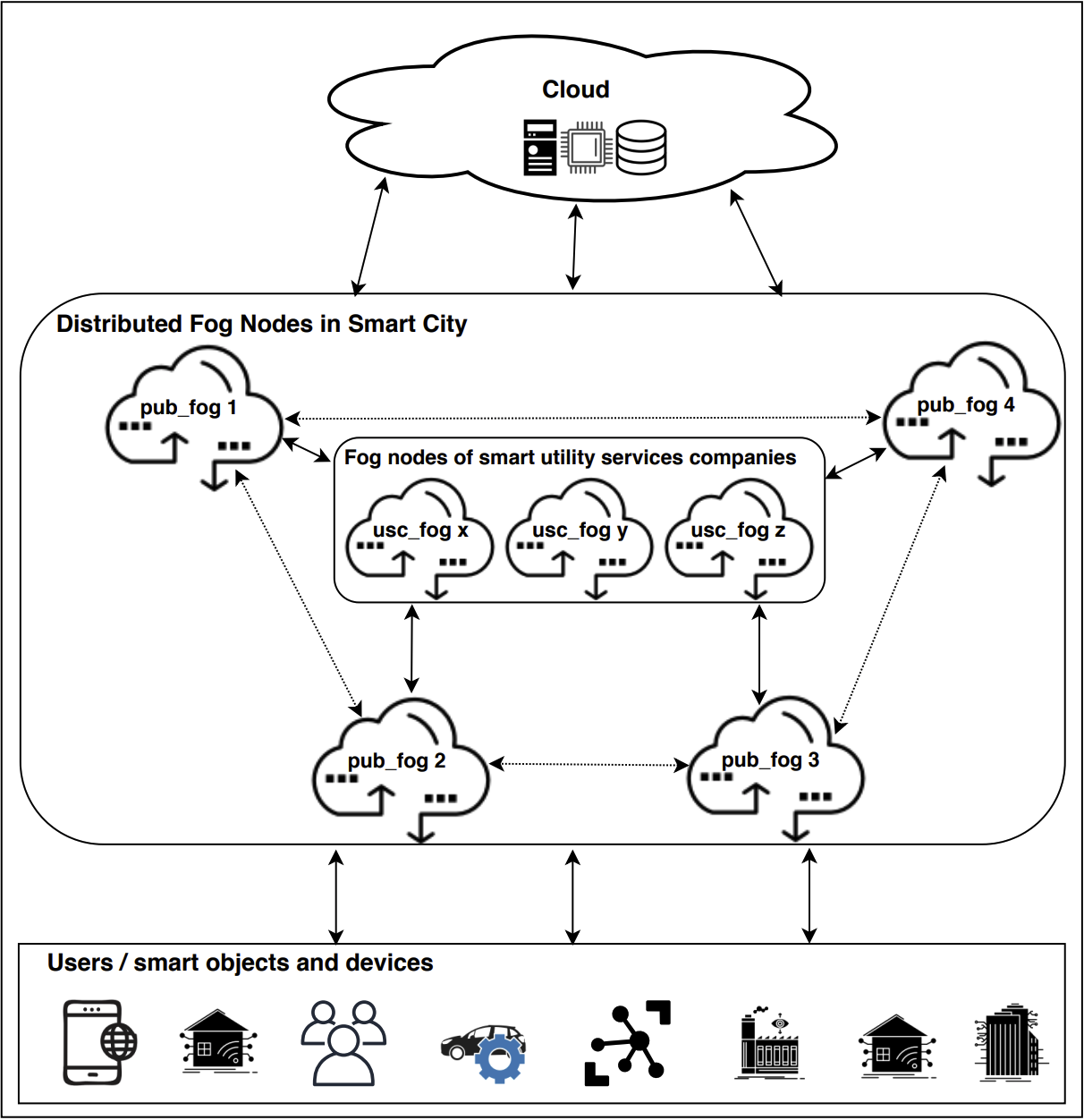}
\caption{The proposed architecture.} \label{ffig1}
\vspace{-0.1 in}
\end{figure}
It is assumed that each pub\_fog node in the system will have the ability to perform some essential data operations, such as calculating customer average usage over a specific time frame or aggregating the totals of selected data values. Those operations are fundamental to optimization of the network bandwidth since the data sent over the network will be diminished by aggregating the necessary data. 
Additionally, as most customer data will be processed locally, at the edge of the network,  it will enhance privacy and security by reducing sensitive data transmittal. Each utility usc\_fog node receiving the data will also perform some essential computations, such as calculating the daily bill and average daily customer usage. These computations by the utilities are important in improving the quality of services in each city as the need for expansion of  services in peak seasons may become evident and shortages avoided. Utilities may use data to plan fuel purchases or raise consumer awareness regarding consumption and conservation. 

\subsection{The Proposed Damage Assessment Algorithms}
\subsubsection{Algorithm 1: The Main Damage Assessment Algorithm}
The IDS is responsible for identifying the attacking transactions and sending a list of them to the victim fog node. Whenever one or more malicious transactions are found on any fog node in the system, the IDS will detect them and send them as a list (MT\_L) to that fog node to be used as input in the proposed schemes. Once the fog node receives the list, it will launch Algorithm 1, which is the main damage assessment algorithm.

As soon as Algorithm 1 is launched, it will create the damage audit table (DA\_Table) and damage data item list (DI\_L). Both will be initialized to null. Then, the algorithm will scan the local log file of the victim fog node, Fog$_1$, beginning from the first attacking transaction of (MT\_L) list, T$_i$. T$_i$ will be added as a new record into DA\_Table since it is the first attacking transaction. If the attacking transaction updates at least one data item, then this data item will be damaged, and any other updating transactions that read this data will be affected as well. It is important to collect and store all data items that have been updated and damaged by the attacking transactions. Then, all transactions that have read those damaged data items can be identified, and data dependency can be declared between the transactions and the fog nodes in the entire system. 

One of the main functionalities of this algorithm will be the collection of data before damage occurs, and store those images, which represent the pre-attack value of the data item, in the written data column on the DA\_Table. These images will be used later in the recovery algorithm. Simultaneously, those damaged data items will be added to the damaged item list to determine data dependency.

Also, the algorithm will examine every transaction in the log file following, the first attack, to determine whether any other transaction is an attacking transaction, or a data access transaction from another fog node, or an updating write transaction. In the case where the transaction is an attacking transaction, the algorithm will perform as in the first attacking transaction. However, if the transaction is an access transaction from another fog node (Fog$_x$), the algorithm will check every data item that has been read by Fog$_x$. A new damage item table for Fog$_x$ (DIT\_Fog$_x$) will be created and all damaged data items that have been read by Fog$_x$ as well as the transaction identification will be added to the DIT\_Fog$_x$. 

Since a fog node may access the same data multiple times, it is essential to know the transaction ID; this will make it easy to find on its log file and confirm that the damaged data items were not corrected later on in the fog$_x$ by valid updating. In the meantime, the DA\_Table will be updated indicating that Fog$_x$ has read the damaged data item, so when the recovery algorithm has successfully corrected the value of the damaged data item, it will send the correct value to Fog$_x$ to use as input for its own recovery algorithm.
\begin{algorithm}[tbp!]
\caption{The Main Damage Assessment Algorithm}
\begin{algorithmic}[1]
\State Create a new DA\_table and initialize to null
\State Create a new DI\_L and initialize to null
\For{every T$_i$ the local log starting from the first attacking transactions of MT\_L} 
    \If {T$_i$ is attacking transaction} 
        \State add it as a new record into DA\_Table 
        \For {every $w_i$ $(A, v1, v2)$}
            \State add $(A, v1)$ pair to data written column
            \State add $A$ to the DI\_L if it is not there
        \EndFor
    \ElsIf{T$_i$ is transaction from another fog node $x$}
        \For {every data item $A$ read by T$_i$ }
            \If{$A$ $\in$ DI\_L}
                \If{DIT\_ Fog$_x$ does not exist}
                    \State Create a new DIT\_ Fog$_x$ where x is the ID of aimed fog node that reads the affected transaction
                    \State Mark T$_i$  as affected transaction
                    \State Add T$_i$ and $A$ into DIT\_ Fog$_x$
                    \State Update the last column of DA\_Table
                \EndIf
            \EndIf
        \EndFor
    \ElsIf{T$_i$ is updating transaction}
        \State add it as a new record into DA\_Table
         \For {every $r_i$ $(A, v)$}
            \If{$A$ $\in$ DI\_L}
               \State add $A$ to invalid read column 
            \Else
                \State add $(A, v)$ to valid read column 
            \EndIf
        \EndFor
        \For {every $w_i$ $(A, v1, v2)$}
            \If {invalid read column of T$_i$ $\neq \varnothing$ }
                \State add $(A, v2)$ to data written column
                \State add $A$ to the DI\_L if it is not there
            \Else { check If ($A$ $\in$ DI\_L)}
                \State add $(A, v2)$ to data written column
                \State delete $A$ from DI\_L
            \EndIf
        \EndFor
        \If{$c_i$ is found \& (both invalid read and data written columns of T$_i$) $= \varnothing$}
            \State delete the record of T$_i$ from DA\_Table
        \ElsIf{$a_i$ is found}
            \State delete the record T$_i$ from DA\_Table  
        \EndIf
    \EndIf
\EndFor
\State Send DIT\_ Fog$_x$ to Fog$_x$ to do further detection
\State Send DA\_Table \& DI\_L for data recovery (algorithm 3) 
\end{algorithmic}
\end{algorithm}
If the transaction is an updating transaction (T$_w$), and not an attacking transaction belonging to the malicious transaction list, then it must be added to the DA\_Table and examined to accomplish two goals. The first goal is to determine data dependency. All read operations must be checked to confirm whether T$_w$ has read any of the damaged data items that already exist to the damaged item list. If so, those damaged data items will be added to the invalid read column of T$_w$, and undamaged data items will be added to the valid read column. Then, all the write operations will be checked to determine whether any have read damaged data items. If so, that means the damage has spread and the written data item is also corrupted. Therefore, it will be added to the damaged item list, if it is not already there.

However, if the transaction T$_w$ updates any data item, (A), without reading any items from the damaged item list, then the data item (A) will be further checked evaluate its inclusion in the damaged item list. If (A) was updated without reading a corrupted transaction, that means it is a valid write and the data item (A) has been refreshed, so (A) must be removed from the damaged item list as in steps(28-30). The new value will be added to the data written column, accomplishing the second goal of adding the non-attacking updating transaction to the DA\_Table.

Finally, all data items in the main victim fog node will be available for use except the damaged data items on DI\_L. Therefore, system availability will be increased. The damage item table DIT\_fog$_x$ will be sent to fog$_x$ to do further detection, while the damage audit table and damage item list will be sent to Algorithm 3, which is the main data recovery algorithm.


\subsubsection{Algorithm 2: Secondary Fog Node Damage Assessment Algorithm}
This algorithm will be like Algorithm 1, with some differences. The main difference in the input of this algorithm is the DIT\_Fog$_x$, which is one of the outputs of Algorithm 1 if an affected fog node reads any damaged data item from the main victim fog node. Assume Fog$_1$ is the main victim fog node in the system, and it was attacked and maliciously updated in the transaction T$_i$, which wrote the data item(Z). Later, Fog$_x$ accessed Fog$_1$ via the transaction T$_j$ to read (Z) and update other data items, (N) and (M), on its database. Here we call Fog$_x$ the secondary affected fog node. Once the secondary affected fog node in our example, Fog$_x$ receives the DIT\_Fog$_x$ containing (Z) as a damaged data item , it will create a new damage audit table and initialize it to null. Note that this algorithm will use the received DIT\_Fog$_x$ to store and track the damaged data items instead of creating a new damage item list. 

\begin{algorithm}[h!]
\caption{Secondary Fog Node Damage Assessment}
\begin{algorithmic}[1]
\Statex Once Fog$_x$ receives the DIT\_ Fog$_x$
\State Create a new DA\_table and initialize to null
\For{every T$_i$ in the local log starting from the first affected transaction on DIT\_ Fog$_x$} 
    \If {T$_i$ $\in$ DIT\_ Fog$_x$ \&\& mark as affected } 
        \State add a record for T$_i$ into DA\_Table 
        \For {every $r_i$ $(A, v)$}
            \If{$A$ $\in$ DIT\_ Fog$_x$}
               \State add $A$ to invalid read column 
            \Else
                \State add $(A, v)$ to valid read column 
            \EndIf
        \EndFor
        \For {every $w_i$ $(A, v1, v2)$}
            \State add $(A, v2)$ pair to data written column
            \State add $A$ into DIT\_ Fog$_x$ if it is not there
        \EndFor
    \ElsIf{T$_i$ is transaction from another fog node $y$}
        \For {every data item $A$ read by T$_i$ }
            \If{$A$ $\in$ DIT\_ Fog$_x$}
                \If{DIT\_ Fog$_y$ does not exist}
                    \State Create a new DIT\_ Fog$_y$ where y is the ID of aimed fog node that reads the affected transaction
                    \State Mark T$_i$  as affected transaction
                    \State Add T$_i$ and $A$ into DIT\_ Fog$_y$
                    \State Update the last column of DA\_Table
                \EndIf
            \EndIf
        \EndFor
    \ElsIf{T$_i$ is updating transaction}
        \State add it as a new record into DA\_Table
         \For {every $r_i$ $(A, v)$}
            \If{$A$ $\in$ DIT\_ Fog$_x$}
               \State add $A$ to invalid read column 
            \Else
                \State add $(A, v)$ to valid read column 
            \EndIf
        \EndFor
        \For {every $w_i$ $(A, v1, v2)$}
            \If {invalid read column of T$_i$ $\neq \varnothing$ }
                \State add $(A, v2)$ to data written column
                \State add a record of T$_i$ into DIT\_ Fog$_x$ with data item $A$ if it is not there
            \Else { check If ($A$ $\in$ DIT\_ Fog$_x$)}
                \State add $(A, v2)$ to data written column
                \State delete $A$ from DIT\_ Fog$_x$
            \EndIf
        \EndFor
        \If{$c_i$ is found \& (both invalid read and data written columns of T$_i$) $= \varnothing$}
            \State delete the record of T$_i$ from DA\_Table
        \ElsIf{$a_i$ is found}
            \State delete the record T$_i$ from DA\_Table  
        \EndIf    
    \EndIf
\EndFor
\State Send DIT\_ Fog$_y$ to Fog$_y$ to do further detection
\State Send DA\_Table \& DIT\_ Fog$_x$ for recovery (algorithm 4) 
\end{algorithmic}
\end{algorithm}

The algorithm will scan the log file and start from the first affected transaction from the received table. Therefore, whenever an affected transaction that belongs to DIT\_fog$_x$ is found, the steps (3-12) will insert it as new record to the damage audit table and check each read operations if its belong to DIT\_fog$_x$ then add it to the invalid read column; otherwise, it will be added to the valid read column. Moreover, for the write operations, the updated data items along with its new values will be added to the data written column as well as they will be added to the DIT\_fog$_x$ table if they are not there. In our example, data items N and M will be added to the DIT\_fog$_x$ and the data written column in DA\_Table. In a like manner, the damage item table, if there is one, will be sent to fog$_y$ while the damage audit table and damage item list will be sent to Algorithm 4, which is data recovery algorithm for the secondary fog node. The process continues until all the affected transactions in the entire system are detected.


\subsection{The Proposed Data Recovery Algorithms}
\subsubsection{Algorithm 3: The Main Data Recovery Algorithm}
Immediately after Algorithm 1 has accomplished its task, it will send the DA\_Table and DI\_L to Algorithm 3 for data recovery. Once Algorithm 3 receives the DA\_Table, it will scan the records that read invalid data items. When a data item record is found in the invalid read column, the algorithm will perform three steps:
\begin{itemize}
	\item { {Step 1:} Scan the data written column upward, beginning at the former record of DA\_Table, for each data item (A) found in the invalid read column. Once the last updated and correct value of (A) is found, this value will be added as a pair (A, v) to the valid read column and data item (A) will be deleted from the invalid read column. This continues until all data items in the invalid read column in the same record have been recovered.}
	\item { {Step 2:} Recompute each data item in the data written column in the same record by reading the new values from the valid read column. Successful completion of Steps 1 and 2 should result in all data items in that record having the correct values.}
	\item { {Step 3:} Check the last column in the same record, which is the fog ID column, to determine if any data item has been read by another fog node in the system; if so, a new Valid Data Items table (VIT\_Fog$_x$) will be created for each affected fog node. Then, the transaction ID along with the correct new value of each accessed data item will be added to the VIT\_Fog$_x$.}
	\end{itemize}
As soon as all the records in the DA\_Table have been examined and all three steps are successfully completed, the VIT\_Fog$_x$ will be sent to the corresponding fog node, Fog$_x$. 

\begin{algorithm}[!tbp]
\caption{The Main Recovery Algorithm}
\begin{algorithmic}[1]

\For{each record in the DA\_Table} 
    \If {invalid read column of T$_i$ $\neq \varnothing$} 
        \For {every data item $A$ in invalid read column}
            \State find the last updated $(A, v)$ pair in data written column of DA\_Table from the former records  
            \State add $(A, v)$ to valid read column
            \State delete $A$ from invalid read column
        \EndFor
        \For {every $A$ in data written column}
            \State  recalculate the value of $A$ using values in the valid read column
        \EndFor
        \If{any fog$_x$ is existing in fog\_ID column}
            \If{ VIT\_ Fog$_x$ does not exist}
                \State Create a new VIT\_ Fog$_x$ where x is the ID of aimed fog node that reads the affected transaction
            \EndIf
            \State Add T$_i$ and the $(A, v)$ pair which is the correct value of $A$ into VIT\_ Fog$_x$
        \EndIf
    \EndIf
\EndFor
\State Send VIT\_ Fog$_x$ to Fog$_x$ node
\For {every $A$ in DI\_L}
    \State check the new log that has just been created while the recovery process was in progress
    \If{$A$ is not modified in the log}
    \State scan data written column of DA\_Table upward to find last updated value of $A$
    \State substitute the value of $A$ in the database with $v$
    \EndIf
\EndFor
\end{algorithmic}
\end{algorithm}

\subsubsection{Algorithm 4: Secondary Fog Node Data Recovery Algorithm}
This algorithm will be like Algorithm 3, with two primary differences. The first is that Algorithm 4 input will be the DA\_Table and the DIT\_Fog$_x$ from Algorithm 2 for the same fog node and the VIT\_Fog$_x$ from another fog node, Fog$_1$. Secondly, this algorithm will check every record on the received DA\_Table. When a transaction that is marked affected is found, then for every data item (A) in the invalid column, the (A, v) pair from the corresponding transaction on the VIT\_Fog$_x$ is copied to the valid data column on DA\_Table and deleted from the invalid column of DA\_Table.
After all the damaged data items have been deleted on each record of the DA\_Table, the data written column will be checked to discern if it is empty. If not, then the value v of each data item (A) in the data written column must be recalculated using the new values in the valid read column. However, the same procedure used in Algorithm 3 will be employed if the record has a transaction with some data items on the invalid read column. The process continues until all affected data items in the system are recovered to a consistent state. 
\begin{algorithm} [!tbp]
\caption{Secondary Fog Node Recovery Algorithm}
\begin{algorithmic}[1]
\Statex Once Fog$_x$ receives the VIT\_ Fog$_x$

\For{each record in the DA\_Table} 
    \If{T$_i$ is affected transaction}
        \For {every data item $A$ in invalid read column}
            \State copy the $(A, v)$ pair from corresponding transaction on the VIT\_ Fog$_x$ to the valid data column on DA\_Table
            \State delete $A$ from invalid read column
        \EndFor
        \If{ the data written column $\neq \varnothing$}
            \For {every $A$ in data written column}
                \State  recalculate the value of $A$ using values in the valid read column
            \EndFor
        \EndIf
     \ElsIf {T$_i$ is updating transaction \& invalid read column $\neq \varnothing$}
        \For{every $A$ in invalid read column}
            \State find the last updated $(A, v)$ pair in data written column of DA\_Table from the former records  
            \State add $(A, v)$ to valid read column
            \State delete $A$ from invalid read column
        \EndFor
        \For {every $A$ in data written column}
            \State  recalculate the value of $A$ using values in the valid read column
        \EndFor
        \If{any fog$_y$ is existing in fog\_ID column}
            \If{ VIT\_ Fog$_y$ does not exist}
                \State Create a new VIT\_ Fog$_y$ where y is the ID of aimed fog node that reads the affected transaction
            \EndIf
            \State Add T$_i$ and the $(A, v)$ pair which is the correct value of $A$ into VIT\_ Fog$_y$
        \EndIf
    \EndIf
\EndFor
\State Send VIT\_ Fog$_y$ to Fog$_y$ node
\For {every $A$ in DIT\_ Fog$_x$}
    \State check the new log that has just been created while the recovery process was in progress
    \If{$A$ is not modified in the log}
    \State scan data written column of DA\_Table upward to find last updated value of $A$
    \State substitute the value of $A$ in the database with $v$
    \EndIf
\EndFor
\end{algorithmic}
\end{algorithm}
\subsection{An Example}
To clarify the proposed scheme, consider the following example. There are two fog nodes in our smart city. Fog$_1$ is a pub\_fog node collecting consumer data via smart meters. Fog$_x$ is a private fog node used by the utility company to manage aggregated data and calculate consumer bills and consumption. Consider the following log schedules for each one of them: 

{\selectfont \textbf{S\textsubscript{Fog.1}}= \textit{r\textsubscript{1}}(\textit{A, 5}) \textit{r\textsubscript{1}}(\textit{B, 4}) \textit{w\textsubscript{1}}(\textit{C, 11, 9}) \textit{w\textsubscript{1}}(\textit{G, 3, 9}) \textit{r\textsubscript{2}}(\textit{B, 4})\textit{c\textsubscript{1 }r\textsubscript{2}}(\textit{G, 9}) \textit{w\textsubscript{2}}(\textit{A, 5, 13}) \textit{w\textsubscript{2}}(\textit{D, 0, 13})\textit{c\textsubscript{2 } fog\textsubscript{x}.r\textsubscript{3}}(\textit{G, 9})\textit{c\textsubscript{3 }w\textsubscript{4}}(\textit{A, 13, 5}) \textit{w\textsubscript{4}}(\textit{G, 9, 3})\textit{c\textsubscript{4} r\textsubscript{5}}(\textit{D, 13}) \textit{r\textsubscript{5}}(\textit{A, 5}) \textit{r\textsubscript{5}}(\textit{C, 9}) \textit{w\textsubscript{5}}(\textit{D, 2, 27})\textit{c\textsubscript{5} r\textsubscript{6}}(\textit{B, 4}) \textit{w\textsubscript{6}}(\textit{B, 4, 4}) \textit{r\textsubscript{6}}(\textit{D, 16}) \textit{w\textsubscript{6}}(\textit{D, 16, 20}) \textit{r\textsubscript{6}}(\textit{A, 5}) \textit{w\textsubscript{6}}(\textit{A, 5, 25})\textit{c\textsubscript{6}} \textit{ fog\textsubscript{x}.r\textsubscript{7}}(\textit{D, 20})\textit{c\textsubscript{7 }r\textsubscript{8}}(C, 9)\textit{c\textsubscript{8 }w\textsubscript{9 }}(C, 9, 11)\textit{c\textsubscript{9} r\textsubscript{10}}(\textit{A, 25})\textit{ r\textsubscript{10}}(\textit{C, 11}) \textit{w\textsubscript{10}}(\textit{E, 10, 36})\textit{c\textsubscript{10}}\textsubscript{\  }\textit{fog\textsubscript{x}.r\textsubscript{11}}(\textit{E, 36})\textit{c\textsubscript{11}}\par}

 \textbf{S\textsubscript{Fog.x}}\textit{ }= \textit{r\textsubscript{9}}(\textit{K, 3) r\textsubscript{9}}(\textit{fog\textsubscript{1}.T\textsubscript{3}.G, 9}) \textit{w\textsubscript{9}}(\textit{K, 3, 12}) \textit{c\textsubscript{9} r\textsubscript{10}}(\textit{M, 10}) \textit{r\textsubscript{10}}(\textit{K, 12}) \textit{w\textsubscript{10}}(\textit{M, 10, 22})\textit{c\textsubscript{10} r\textsubscript{14}}(\textit{fog\textsubscript{1}.T\textsubscript{7}.D, 20}) \textit{r\textsubscript{14}}(\textit{L, 4}) \textit{w\textsubscript{14}}(\textit{N, 17, 24}) \textit{c\textsubscript{14\  }r\textsubscript{16}}(\textit{fog\textsubscript{1}.T\textsubscript{11.}E, 36}) \textit{w\textsubscript{16}}(\textit{P, 4 , 36})\textit{c\textsubscript{16}}

Now, suppose the IDS detects the first transaction, T$_1$, on the Fog$_1$ schedule is an attacking transaction and data items (C) and (G) are detected as having been maliciously updated. The IDS will send T$_1$ as the list MT\_L to Fog$_1$. Once Fog$_1$, the primary victim fog node in the system, receives the list, it will launch Algorithm 1. Then, it will create a new DA\_Table and a new DI\_L. Consequently, the log file of Fog$_1$ will be scanned, beginning with the first attacking transaction on the MT\_L, which is T$_1$. 

Whenever an attacking transaction is found, such as T$_1$ in this example, it will be added to the DA\_Table as a new record. All the write operations of T$_1$ will also be checked, so whenever a data item is found, it will be added along with its old value (before image) as a pair to the data written column, and the data items will be added to the damaged items list. In our example, the pairs (C, 11) and (G, 3) are added to the data written column (Table~\ref{table3}) while the data items (C) and (G) will be added to DI\_L. This will be the case for all attacking transactions that belong to MT\_L.

The algorithm will examine the next transaction in the log file, which is T$_2$. Since T$_2$ is an updating transaction, it will be added into the DA\_Table. Consequently, every reading operation in T$_2$ will be examined to ascertain if it read any damaged data item from the DI\_L; if so, the data item will be added to the invalid read column, as apparent with (G). Otherwise, it will be added, as a pair with its value, to the valid read column as (B, 4) in the example. The next transaction is T$_3$. Fog$_x$ has read the data item (G), which is a damaged data item. So, the Fog ID column will be marked and a new Damage Items Table will be created for Fog$_x$ (DIT\_Fog$_x$), adding data item (G) to the table with the transaction ID (Table~\ref{table4}).

The algorithm will also find that damaged data items (A) and (G) have been refreshed in T$_4$ and updated without reading any other damaged data items. Therefore, they will be added, with their new values, to the data that were written and removed from the DI\_L. The process continues until the end of the log is reached. Then, the DIT\_Fog$_x$ will be sent to Fog$_x$ to be used as input for Algorithm 2, while the DA\_Table and DI\_L will be sent to Algorithm 3, which is the primary recovery algorithm. All data items on DI\_L will be blocked from being used until they recover.

Once Fog$_x$ receives the DIT\_Fog$_x$, it will launch Algorithm 2 and use the DIT\_Fog$_x$ as input in further assessment and detection processes. Note that this table will add any damaged data items that are detected in Fog$_x$, ( Table~\ref{table7}). Thus, a new DA\_Table will be created. It will scan the local log of Fog$_x$, starting from the first affected transaction on the DIT\_Fog$_x$, which is T$_9$. T$_9$ will be added to the new DA\_Table and for every read operation, the data item will be examined to find out if it belongs to the DIT\_Fog$_x$; if so, it will be added to the invalid read column. Otherwise, it will be added to the valid read column along with its value. Therefore, the data item fog$_1$.T$_3$.G will be added to the invalid read column, while the pair (K, 3) will be added to the valid read column (Table~\ref{table5}). However, it will do the same thing for the write operations as in Algorithm 1, so the updated data item (K) along with its value (K, 12) will be added to the data written column. 

Meanwhile, the data item (K), will be added into the DIT\_Fog$_2$ (Table~\ref{table7}), since it becomes affected by reading the damaged data item fog$_1$.T$_3$.G. For T$_{10}$, the updating occurs after reading the damaged data item (K), so the scenario (the same as shown in T$_2$ in Fog$_1$) will be repeated (Table~\ref{table5}). Consequently, the process continues until the end of the log is reached. Once reached, the DA\_Table for Fog$_x$ will be sent to Algorithm 4 to conduct data recovery. And all data items on Table~\ref{table7} will be blocked until they recover.

As soon as Fog$_1$ has completed the damage assessment algorithm, Algorithm 1, and sent the DA\_Table and DI\_L to Algorithm 3 to proceed with data recovery, which will be launched immediately, and received the DA\_Table and DI\_L as inputs, it will scan the DA\_Table from its beginning and search for any transactions that read invalid data items. For example, T$_2$ read invalid data item (G), and the algorithm looks for the last valid update value of (G), which must be the closest transaction before T$_2$. Therefore, T$_1$ must have the latest updated correct value of (G), which is (3). The pair (G, 3) will be copied to the valid read column, and (G) will be removed from the invalid read column. After that, T$_2$ will be recalculated using the new values (Table~\ref{table10}). (Note that in this example any transaction where write operations are found after read operations, all values of the read operations will be added together.) 

Following T$_2$, T$_3$ will be processed in the same manner. As Fog$_x$ has read the damaged data item (G), a new Valid Data Item Table for Fog$_x$ (VIT\_Fog$_x$) will be created and added to the transaction ID, T$_3$, and the correct value of (G), which is (G, 3) will be added (Table~\ref{table11}), and so on until the end of the DA\_Table. After that, VIT\_Fog$_x$ will be sent to Fog$_x$ to be used as an input in Algorithm 4 to recover the data.

Once Fog$_x$ receives the VIT\_Fog$_x$, it will launch Algorithm 4 and use VIT\_Fog$_x$ along with it is own DA\_Table to process data recovery. Then, every record in the DA\_Table will be checked. Since the first record, T$_9$ in this example, must be an affected transaction from Fog$_1$, then VIT\_Fog$_x$ should have the correct and valid value of the damaged data item. Therefore, the new value of data item (G) will be copied from VIT\_Fog$_x$ to the valid read column of the DA\_Table and removed from the invalid read column. After that, T$_9$ will be recalculated using the new values (Table~\ref{table12}). The rest of the algorithm will be almost the same as Algorithm 3.






\begin{table}[htbp]
\caption{The Damage Audit Table for Fog$_1$ }
\begin{center}
\begin{tabular}{|c|c|c|c|c|}
\hline
\textbf{T Id}&\textbf{Data written}&\textbf{Valid read}&\textbf{Invalid}&\textbf{Fog ID}\\
\hline
T$_1$ & (C, 11), (G, 3) &   & & \\
\hline
T$_2$ & (A, 13), (D, 13)  & (B, 4)   & G & \\
\hline
T$_3$ &   &    & G & Fog$_x$ \\
\hline
T$_4$ & (A, 5), (G, 3)   &    &  &  \\
\hline
T$_5$ & (D, 27)   & (A, 5)   & C, D &  \\
\hline
T$_6$ & (D, 20), (A, 25)   & (B, 4), (A, 5)   & D &  \\
\hline
T$_7$ &    &    & D & Fog$_x$  \\
\hline
T$_9$ & (C, 11)   &    &  &   \\
\hline
T$_{10}$ & (E, 36)    & (C, 11)   & A  &   \\
\hline
T$_{11}$ &     &    & E  & Fog$_x$  \\
\hline
\end{tabular}
\label{table3}
\end{center}
\vspace{-0.2 in}

\end{table}

\begin{table}[htbp]
\caption{Fog$_x$ Damage Item Table Created by Fog$_1$}
\begin{center}
\begin{tabular}{|c|c|}
\hline
\textbf{Transaction Id}&\textbf{Damaged Data Items}\\
\hline
fog$_1$.T$_3$    &   G   \\
\hline
fog$_1$.T$_7$ &   D  \\
\hline
fog$_1$.T$_{11}$  &   E   \\
\hline
\end{tabular}
\label{table4}
\end{center}
\vspace{-0.2 in}
\end{table}


\begin{table}[htbp]
\caption{The Damage Audit Table for fog$_x$ }
\begin{center}
\begin{tabular}{|c|c|c|c|c|}
\hline
\textbf{T Id}&\textbf{Data written}&\textbf{Valid read}&\textbf{Invalid}&\textbf{Fog ID}\\
\hline
T$_9$ & (K, 12) & (K, 3)  & fog$_1$.T$_3$.G & \\
\hline
T$_{10}$ & (M, 22)   & (M, 10)   & K & \\
\hline

T$_{14}$ & (N, 24)   & (L, 4)    & fog$_1$.T$_7$.D  &  \\
\hline
T$_{16}$ & (P, 36)    &    & fog$_1$.T$_{11}$.E &  \\
\hline
\end{tabular}
\label{table5}
\end{center}
\vspace{-0.2 in}
\end{table}

\begin{table}[htbp]
\caption{DIT\_Fog$_x$ with all damaged data items that are found on Fog$_x$}
\begin{center}
\begin{tabular}{|c|c|}
\hline
\textbf{Transaction Id}&\textbf{Damaged Data Items}\\
\hline
fog$_1$.T$_3$    &   G     \\
\hline
fog$_1$.T$_7$ &   D   \\
\hline
fog$_1$.T$_{11}$  &   E    \\
\hline
T$_9$ &   K   \\

\hline
T$_{10}$ &   M   \\

\hline
T$_{14}$ &   N   \\
\hline
T$_{16}$ &   P  \\
\hline
\end{tabular}
\label{table7}
\end{center}
\vspace{-0.2 in}

\end{table}
\begin{table}[!htbp]
\caption{DA\_Table for fog$_1$ after damaged data have been recovered}
\begin{center}
\begin{tabular}{|c|c|c|c|c|}
\hline
\textbf{T Id}&\textbf{Data written}&\textbf{Valid read}&\textbf{Invalid}&\textbf{Fog ID}\\
\hline
T$_1$ & (C, 11), (G, 3) &   & & \\
\hline
T$_2$ & (A, 7), (D, 7)  & (B, 4), (G, 3)   &  & \\
\hline
T$_3$ &   &  (G, 3)  &  & Fog$_x$ \\
\hline
T$_4$ & (A, 5), (G, 3)   &    &  &  \\
\hline
T$_5$ & (D, 23)   & (A,5), (C, 11), (D, 7)  &  &  \\
\hline
T$_6$ & (D, 27), (A, 32)   &(B, 4), (A, 5), (D, 23)   &  &  \\
\hline
T$_7$ &    & (D, 27)   &  & Fog$_x$  \\
\hline
T$_9$ & (C, 11)   &    &  &   \\
\hline
T$_{10}$ & (E, 43)    & (C, 11), (A, 32)   &   &   \\
\hline
T$_{11}$ &     & (E, 43)   &   & Fog$_x$  \\
\hline
\end{tabular}
\label{table10}
\end{center}
\vspace{-0.2 in}

\end{table}

\begin{table}[htbp]
\caption{VIT\_Fog$_x$ sent from fog$_1$}
\begin{center}
\begin{tabular}{|c|c|}
\hline
\textbf{Transaction Id}&\textbf{Valid Data Items}\\
\hline
fog$_1$.T$_{3}$ &  (G,3) \\
\hline
fog$_1$.T$_{7}$ &   (D,27) \\
\hline
fog$_1$.T$_{11}$ &    (E,43) \\
\hline
\end{tabular}
\label{table11}
\end{center}
\vspace{-0.2 in}

\end{table}

\begin{table}[!htbp]
\caption{DA\_Table for fog$_x$ after damaged data have been recovered}
\begin{center}
\begin{tabular}{|c|c|c|c|c|}
\hline
\textbf{T Id}&\textbf{Data Written }&\textbf{Valid Read}&\textbf{Invalid}&\textbf{fog$_{ID}$}\\
\hline
T$_{9}$    &   (K, 6)    &   (K,3), (fog$_1$.T$_3$.G, 3) &  &  \\
\hline
T$_{10}$ &  (M, 16)  &   (M, 10), (K, 6) & & \\
\hline
T$_{14}$ &   (N, 31)  &   (L, 4), (fog$_1$.T$_7$.D, 27)&  &  \\
\hline
T$_{16}$ &   (P, 43)  &   (fog$_1$.T$_{11}$.E, 43) & & \\
\hline
\end{tabular}
\label{table12}
\end{center}
\vspace{-0.2 in}

\end{table}

%
\newpage
\section {conclusion}
Intrusion detection is one of the main phases that must be included to ensure the security and reliability of any computing system. This phase uses software or device to observe the system for any malicious activity or policy violation. However, detection systems sometimes fail to detect several malicious transactions on time, leading to data damage. Therefore, intrusion detection must be complemented by another phase, namely, damage assessment and data recovery, which ensures the integrity and availability of system data. This phase identifies any further affected transactions and ensures that the database returns to a consistent state. In this paper, we have introduced a novel architecture model for applying fog technology to smart cities. Working with the nature and characteristics of the model, we propose a unique method of assessing and recovering damaged data. As part of future work, we plan to evaluate the proposed model by simulating the whole environment and examining the performance of our algorithms.

\end{document}